# Periodic DOS modulation in underdoped Bi2223

Choichiro Nakashima[a*], Tetsuya Iye[b], Azusa Matsuda[a]

[a]*Department of Physics, School of Science and Engineering, Waseda University, Okubo 3-4-1, Shinjuku, Tokyo 169-8555, Japan*
[b]*Department of Physics, Graduate School of Science, Kyoto University, Kyoto 606-8502, Japan*

**Abstract**

In this report, we show the results of the STM/STS study on the $Bi_2Sr_2Ca_2Cu_3O_{10+\delta}$ (Bi2223) single crystals whose doping levels were ranging from an optimally doped to an underdoped region. Bi2223 single crystals were grown by a TSFZ method and their doping levels were adjusted with annealing in an oxygen deficient atmosphere. Single crystals were cleaved in an UHV at around 77 K and STM/STS measurements were carried out in the same conditions. We successfully obtained the tunneling spectrum maps as well as topographic images. We found that the superconducting gap was much more homogeneous than in the case of the Bi2212 in optimal doping, but becomes inhomogeneous with decreasing a doping level. This suggests the decoupling of the three Cu-O layers in terms of the SC correlation. More importantly, we found a new periodic modulation in the LDOS map with periods about $2a_0$, which was almost dispersion-less and observed only in the underdoped $T_C$=85 K sample. This modulation is possibly related to the charge/spin order in the inner plane, which is supposed to be highly undrerdoped.

## 1. Introduction

Multilayered high-$T_C$ superconductors have been studied for a long time. One of the interests is that multilayering enhances superconducting transition temperature ($T_c$). In this respect, it is important to study the interaction between the Cu-O layers. However, because of the difficulty in preparing the high quality single crystals of the multilayered superconductors, not many studies have been done to clarify the characteristic of the interaction. Another interesting point in the multilayered superconductors is the possible co-existence of a superconducting and an antiferromagnetic (AF) Cu-O layers, due to inhomogeneous carrier doping among the layers. That was believed to be realized in the five-layer superconductor $HgBa_2Ca_4Cu_5O_{12+\delta}$ and $Ba_2Ca_4Cu_5O_{10}(F,O)_2$ [1,2]. However, the actual existence of the co-existence phase in the other materials and the superconducting properties of a co-existence phase have not been studied well. Here, we report the doping dependence of the superconducting characteristics of the three-layered superconductor $Bi_2Sr_2Ca_2Cu_3O_{10+\delta}$ (Bi2223), focusing on the interaction between Cu-O layers, using a low temperature scanning tunneling microscope (STM).

The STM has an ability to map a spatially inhomogeneous electronic system and has revealed many interesting spatial modulations in a $Bi_2Sr_2CaCu_2O_{8+\delta}$ (B2212) superconductor, such as inhomogeneous gap distribution [3,4], a quasi-particle interference pattern [5], a $4a_0$ modulation [6,7] and a nematic order [8]. Except for quasi-particle interference, their origins were not well understood. As long as the mechanisms of inhomogeneity are concerned, it is also interesting to study the effect of multilayering or the interaction between the adjacent Cu-O layers. In Bi2223, we found that dimensionality of superconductivity can be changed by a carrier concentration [9], indicating that it is possible to adjust the strength of the superconducting coupling. We expect the study of inhomogeneous states in Bi2223 gives valuable information on a possible electric/magnetic order in the inner plane (IP) and is helpful to identify the origin of the various inhomogeneous states observed in Bi2212.

## 2. Experiment

The single crystals were grown by a travelling solvent floating zone method with the condition described in ref. [9,10]. The sintered rods for single crystal growth were prepared with a cation ratio of Bi:Sr:Ca:Cu =2.1:1.9:2:3. Obtained single crystals were annealed in an atmosphere with controlled partial oxygen pressures. We show the results of samples annealed in (1) 760 torr $O_2$ at 600℃, and (2) $7.6\times10^{-5}$ torr $O_2$ at 590℃. $T_c$s, which were determined by DC susceptibility measurements, were 108 K and 85 K, respectively. They corresponded to a nearly optimally doped and an underdoped state, respectively. The samples were in-situ cleaved at 78 K in $10^{-10}$ torr and STM/STS measurements were carried out with a same temperature and a vacuum condition. W tips, prepared by the electrochemical etching in a KOH solution, were used. All STM images were taken by a constant height mode with a sample bias at 300 mV and a resolution at 512×512. The spectral maps were acquired by taking IV characteristics at each 128×128 data points with their biases varying from -300 mV to 300 mV. We could obtain a topographical image and a spectral map at an approximately same area.

---

* Choichiro Nakashima. Tel.: +81-3-5286-3511; fax: +81-3-5286-3511.
*E-mail address*: cho-nakashima@asagi.waseda.jp.

## 3. Result and Discussion

Fig.1 a) and c) show topographic images of the sample (1) ($T_c$=108 K) and the sample (2) ($T_c$=85 K), respectively. We can see a super lattice modulation typical to Bi2223 and Bi2212. Fig.1 b) and d) are the gap maps of samples (1) and (2). Fig.1 f) shows the dI/dV spectra taken along the red line shown in Fig.1 a) and c). Fig.1 b) and f) show that a superconducting gap of the optimally doped sample (1) is almost uniform, contrary to the case in Bi2212 [3,4], where apparent gap distribution has been observed. Note that because the measurement was done at 78 K, the spectra were thermally broadened. Nevertheless, we can see well-defined gaps at about 60 mV, which are typical to the Bi2223 samples. However, once the carrier is reduced to an underdoped level, the gap becomes inhomogeneous as shown in Fig.1 d). This is reasonably understood if we consider the change in dimensionality of superconductivity specific to a multilayered Bi2223. In the reference [9], we investigated a superconducting fluctuation effect and found that the underdoped Bi2223 behaves like two-dimensional superconductor similar to Bi2212. However, the increase in the doping level alters superconductivity more three dimensional in Bi2223, although Bi2212 stays in a 2d superconductor irrespective to its doping level. This is understood as an effect of inhomogeneous doping in the multilayered superconductor. Bi2223 has two nonequivalent layers, two outer pyramidal Cu-O planes (OP) and one planer inner plane (IP). If the IP and the OP are coherent in terms of superconductivity, they are expected to behave an anisotropic three-dimensional superconductor. However, if the coherence is lost due to the loss of superconductivity in the strongly underdoped IP, it becomes 2d like. On the other hand, there is some evidence that the source of gap inhomogeneity comes from outside the Cu-O planes, possibly the interstitially introduced excess oxygen atoms in a Bi-O plane [11]. If this is a case, the effect to superconductivity should be strongly reduced when superconductivity acquires three-dimensionality, which might be realized in the optimally doped Bi2223.

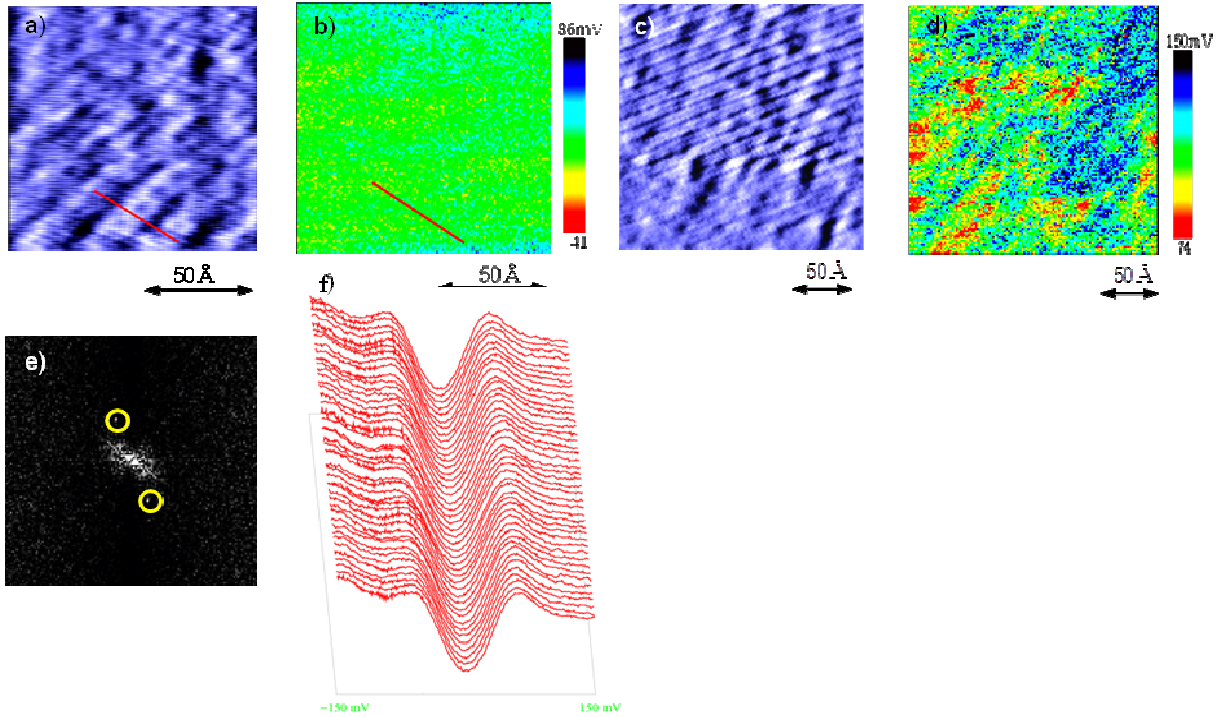

Fig.1. a) A topographic image, and b) a gap map for the $T_c$=108 K sample. The area sizes are 100x100 Å; c) A topographic image, and d) a gap map for $T_c$=85 K sample. The area sizes are 200x200 Å; e) The FT image of d). The yellow circles coincide to the period due to the Bi-O modulation; f) Spectra taken along the red lines shown in a) and b).

To see the structure observed in Fig.1 d) in detail, the Fourier transformed image was shown in Fig.1 e). We can see a periodic structure shown by the yellow circles. The period corresponds to the super lattice modulation in a Bi-O plane, and is consistent with the previous observation [12].

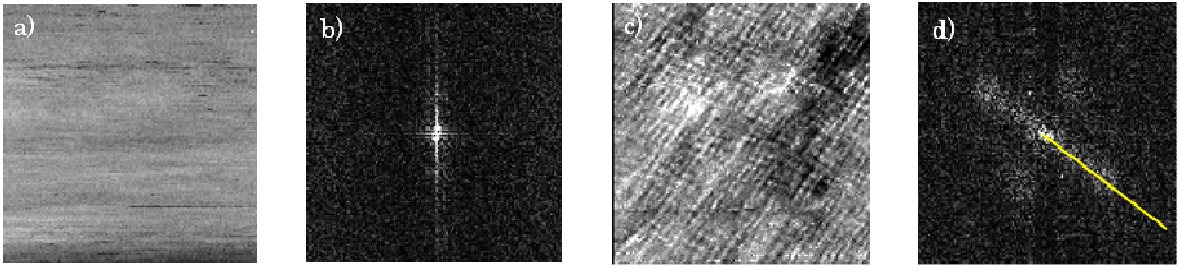

Fig.2. a) a DOS map of a sample (1) at -10 mV; b) the FT image of a); c) a DOS map of a sample (2) at -6 mV; d) the FT image of c)

Next, we calculated DOS maps at fixed bias voltages from a set of I-V characteristics, in order to search the spatial modulation of the electronic system. The obtained DOS map at -10 mV of a sample (1) was shown in Fig.2 a) and that at -6 mV of a sample (2) was shown in Fig.2 c). Their FT images were shown in Fig.2 b) and d), respectively. As is evident from Fig.2 a) and b), the optimally doped sample has a uniform electronic system. On the other hand, in Fig.2 c) the DOS is apparently non-uniform and has several characteristic structures. We observe (a) a periodic structure with a period of about $2a_0$ along the Cu-O-Cu bonding directions, (b) a non periodic background structure, which has some correlation with the topographic protrusions and recessions seen in Fig.1 c). The structures of a type (b) are commonly observed also in Bi2212. On the other hand, the $2a_0$ periodic structure (a) is specific to Bi2223 and reported for the first time.

To see the energy dependence of the modulation pattern, DOS maps at -18 mV and 5 mV were shown in Fig.3 a) and c), as well as their FT maps in Fig.3 b) and d). To compare these maps more quantitatively, the intensity of the FT maps along the line shown in the Fig.2 d) were plotted in Fig. 3 e) for several bias voltages. Fig. 3e) shows that the observed $2a_0$ DOS modulation is independent of the quasi-particle energy, that is, non dispersive. Therefore it is not related to quasi-particle interference.

There are several scenarios to explain this new $2a_0$ modulation. First of all, we need to consider which plane gives the observed pattern. Since the ARPES and NMR study of multilayered superconductor [13,14] indicates the OPs are highly overdoped even in the optimally doped sample, we think the OPs had stayed overdoped both in a sample (1) and (2). Then, it is hard to consider that the pattern comes from OP. Then, we are probing a modulation in IP through the interaction between IP and OP, which can be electronic or magnetic. Since our anisotropy measurements indicate the decoupling of the superconducting correlation in a sample (2), the IP should be non-superconducting in a sample (2). The simplest assumption is that IP becomes an AF state in (2). However, this probably gives a 45° rotated $\sqrt{2}a_0$ modulation, which contradicts our observation. Therefore, if the spin ordering is an origin of this modulation, there should be a stripe or a SDW formation. In this case, we are looking at a superconductor coexisting with a spin order. On the other hand, the existence of a $4a_0$ modulation has been reported in Bi2212, where the origin is supposed to be a CDW formation due to the Fermi surface nesting near the anti-node regions. We may consider that the $2a_0$ modulation also comes from a CDW formation. Since the IP might be highly underdoped, the period can be shortened to $2a_0$. In this case, there exists a rather strong electronic interaction between IP and OP, even in an underdoped region. To specify the mechanism of the modulation, further studies are required. Especially, doping dependence measurements may be effective to discriminate what is happening.

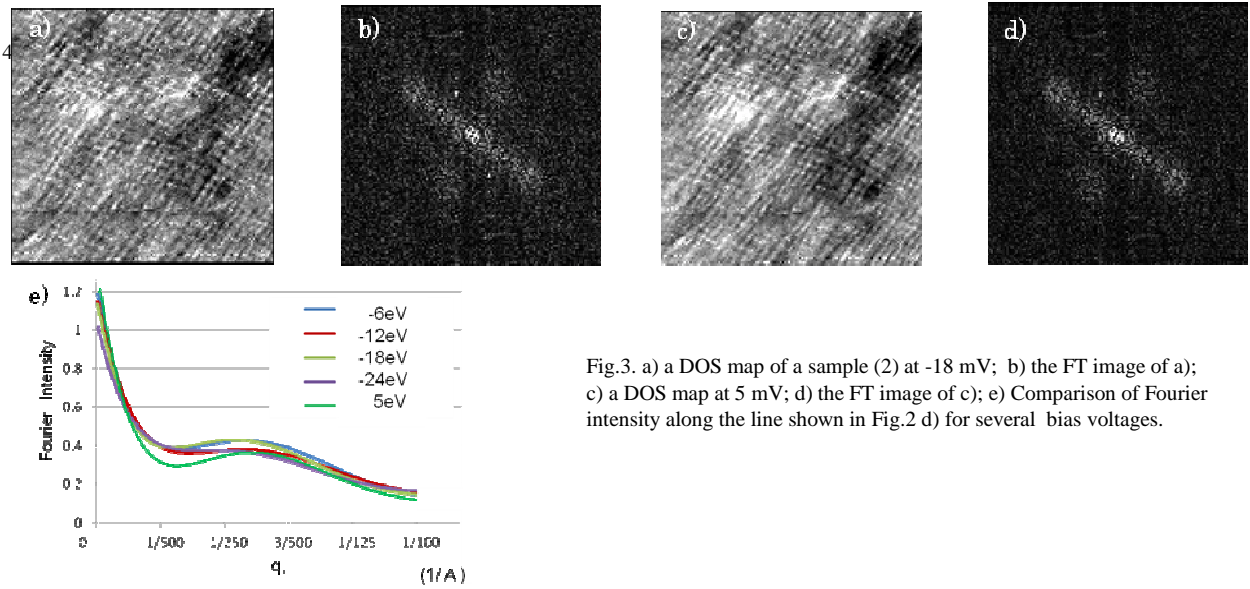

Fig.3. a) a DOS map of a sample (2) at -18 mV;  b) the FT image of a); c) a DOS map at 5 mV; d) the FT image of c); e) Comparison of Fourier intensity along the line shown in Fig.2 d) for several bias voltages.

## 4. Summary

We have studied the electronic systems of Bi2223 single crystals of $T_c$=108 K (optimally doped) and of $T_c$=85 K (underdoped) by an LT-STM. We found that in the optimally doped samples a spatially uniform gap was observed, while underdoping derives it non-uniform. This is consistent with the result of the dimensionality analysis using a superconducting fluctuation, where overdoping derives the system more 3d-like in terms of superconductivity in Bi2223. More importantly, we found a new non-dispersive modulation structure with a period $2a_0$ in the DOS of the underdoped samples. This structure is possibly come from an unknown electronic order in the highly underdoped inner plane and could be transferred to the outer planes through electronic interaction between the planes. Further study may reveal the characteristics of the order and the interaction between planes.